\documentstyle[12pt, epsfig]{article}
\begin{document}

\begin{titlepage}

\title{ An exact solution on the ferromagnetic Face-Cubic spin model on a Bethe lattice.}

\author{ V. R. Ohanyan$^{1,2}$ , L. N. Ananikyan$^1$, N. S. Ananikian$^1$ \\[1mm]
{\small \sl $^1$ Department of Theoretical Physics, Yerevan Physics Institute,} \\
{\small \sl Alikhanian Brothers 2, 375036 Yerevan, Armenia}\\[1mm]
{\small \sl $^2$ Chair of Theoretical Physics, Yerevan State University,} \\
{\small \sl Al. Manoogian 1, 375049, Yerevan, Armenia}
 \\[1mm]}

\maketitle

\begin{abstract}
The lattice spin model with $Q$--component discrete spin variables
restricted to have orientations orthogonal to the faces of
$Q$-dimensional hypercube is considered on the Bethe lattice, the
recursive graph which contains no cycles. The partition function of
the model with dipole--dipole and quadrupole--quadrupole interaction
for arbitrary planar graph is presented in terms of double graph
expansions. The latter is calculated exactly in case of trees. The
system of two recurrent relations which allows to calculate all
thermodynamic characteristics of the model is obtained. The
correspondence between thermodynamic phases and different types of
fixed points of the RR is established. Using the technique of simple
iterations the plots of the zero field magnetization and quadrupolar
moment are obtained. Analyzing the regions of stability of different
types of fixed points of the system of recurrent relations the phase
diagrams of the model are plotted. For $Q \leq 2$ the phase diagram
of the model is found to have three tricritical points, whereas for
$Q> 2$ there are one triple and one tricritical points.

 PACS number(s): 05.50.+q, 05.70.Fh
\end{abstract}
\end{titlepage}

\section{Introduction}
 Cubic symmetry plays a  significant role in many types of phase
 transitions and critical phenomena. Still in 70-s various models
possessing cubic symmetry have been introduced to determine the
nature of the displacive phase transitions in perovskites
\cite{aha74,bru75}. Magnetic phenomena in cubic crystals are also
affected by the lattice structure.
  For instance, in the crystalline solids
 with cubic--symmetric lattices (Fe, Ni, etc.) it leads to the
 modification of the magnetic exchange interaction giving rise to
 additional contributions to the conventional O(N)-symmetric Heisenberg Hamiltonian.
 The simplest contribution of the underlying cubic symmetry is the
 single-ion anisotropy of the form ${\sum}_i \mathbf{s}_i^4$.
 During the last three decades different aspects of this
 issue have been the subject of various investigations.

    In the framework of the field--theoretical approach to critical
phenomena modelling of effects of cubic symmetry is usually
performed in
 terms of continuous-spin Landau-Ginzburg effective Hamiltonian with cubic
 anisotropy, i.e. the $\phi^4$ - theory with an additional
 cubic term which breaks explicitly the O(N) invariance to a residual discrete cubic symmetry \cite{aha76} - \cite{cal04}:
 \begin{eqnarray}
 - \beta{\mathcal{H}}= \int d^Dx
 \left\{\frac{1}{2}\left(\partial_{\mu} \phi \right)^2+
 \frac{1}{2}a\phi^2+\frac{1}{4}b\phi^4+
 v \sum_{\alpha=1}^{N} (\phi_{\alpha})^4\right\}, \label{1}
 \end{eqnarray}

 where $\phi(x)= (\phi_1,\phi_2,\ldots ,\phi_N) $ is a continuous N-component local vector order
 parameter, $\phi^4 = (\phi^2)^2$ and $\beta=1/kT$ is inverse temperature.
 The spin orientations, orthogonal to faces of an $N$-dimensional
 hypercube will be favorable for $v<0$, whereas positive values of
 $v$ favor the orientations toward the corners.

    This model received numerous important applications, among
    which are such as the oxygen ordering in YBaCu$_3$O$_{6+x}$
    \cite{bar89}, one of the most studied high-T$_c$
    superconductor; the buckling instability of confined colloid
    crystal layer \cite{cho93}; the micellar binary solution of
    water and amphiphile \cite{bek00} etc. The two--dimensional
    case has been recently intensively studied within the framework
    of renormalization group technique in the space of fixed
    dimensionality up to five--loop approximation
    \cite{cal02}-\cite{cal04}. The model was found to have four
    fixed points: the Gaussian one, the Ising one with N decoupled
    components, the O(N)--symmetric and the cubic fixed point.
    Analogous calculations were made for $D=3$, as well revealing
    some peculiar type of cubic fixed point for $N>2$
    corresponding to specific anisotropic mode of the critical
    behavior \cite{kle97}-\cite{cal02a}. The surface critical
    behavior of the model was also studied \cite{usa04}. It has been
    recently shown that the behavior of the spherical many--spin
    magnetic nanoparticle with surface anisotropy may be modelled
    by an effective single macro--spin with cubic anisotropy terms
    in the effective energy \cite{kac06}.

        Another class of lattice cubic--symmetric spin models was
        introduced by Kim, Levy and Uffer in Ref. \cite{kim75} to
        explain the tricritical--like behavior of cubic rare--earth
        compounds, particularly holmium antimonide, HoSb
        \cite{kim75}-\cite{nie83}. Projecting the pair exchange
        interaction onto the sixfold degenerated ground-state
        manifold of Ho$^{3+}$ ion they arrived at the following
        effective Hamiltonian:
\begin{eqnarray}
-\beta {\mathcal{H}} = \sum_{\langle
 i,j\rangle}J_{i  j}{\mathbf{S}}_i{\mathbf{S}}_j, \label{1.2}
\end{eqnarray}
where the classical spin variables ${\mathbf{S}}_i$ are the unit
vectors restricted to have orientations, orthogonal to the faces of
the cube and the sum is going over all the pairs of
nearest--neighbor sites. Generalization for $Q$ component spin is
obvious: each spin can assume $2Q$ orientations:
\begin{eqnarray}
{\mathbf{S}}_i \in \{(\pm1, 0, \ldots, 0), (0, \pm1, \ldots, 0)
\ldots (0, \ldots, 0, \pm1)\}. \label{1.3}
\end{eqnarray}
This model is known as the Face-cubic model. It is connected to the
continuous cubic model of Eq. (\ref{1}) via the limit of strong
anisotropy $(|v|\gg |b|)$. In a similar way one can also consider
 quadrupolar pair interaction terms and obtain the following
Hamiltonian provided all interactions are homogenous \cite{kim76,
aha77}:
\begin{eqnarray}
-\beta {\mathcal{H}}_{FC}=J\sum_{\langle i,j\rangle}
({\mathbf{S}}_i{\mathbf{S}}_j)+K\sum_{\langle i,j\rangle}
({\mathbf{S}}_i{\mathbf{S}}_j)^2. \label{1.4}
\end{eqnarray}
It is easy to see that Hamiltonian (\ref{1.4}) can be represented in
terms of two sets of discrete variables: a Potts--like determining
which component of ${\mathbf{S}}_i$ is non-zero and an Ising--like,
corresponding to the sign of the component. Indeed,
\begin{eqnarray}
{\mathbf{S}}_i{\mathbf{S}}_j=\sigma_i \sigma_j \delta_{\alpha_i
\alpha_j}, \label{1.5}
\end{eqnarray}
where $\sigma_i = \pm1$ and $\alpha_i = 1,2,...Q$. Therefore we have
 \begin{eqnarray}
 - \beta{\mathcal{H}}_{FC}=J\sum_{\langle i,j\rangle}\sigma_i \sigma_j
 \delta_{\alpha_i, \alpha_j}+K\sum_{\langle i,j\rangle }\delta_{\alpha_i,
 \alpha_j}. \label{1.6}
\end{eqnarray}
Formally we can enlarge the Hamiltonian (\ref{1.4}) up to the
interaction terms of higher power of the
$({\mathbf{S}}_i{\mathbf{S}}_j)$:
\begin{eqnarray}
-\beta {\mathcal{H}}_{FC}^{(L)}=\sum_{\langle i,j\rangle}
\sum_{n=0}^L J_n ({\mathbf{S}}_i{\mathbf{S}}_j)^n .\label{1.7}
\end{eqnarray}
For arbitrary finite $L$ this expression leads to the same
Hamiltonian that of Eq. (\ref{1.6}) with $J=\sum_{k=0}^L J_{2k+1}$
and $K=\sum_{k=0}^L J_{2k}$. At $K=0$ the Hamiltonian (\ref{1.7})
reduced to the $2Q$ -- state Potts model; $J=0$ corresponds to two
decoupled $Q$ -- state Potts model, and at $Q=2$ one obtains the
Ashkin--Teller model. Variational renormalization--group study of
the pure and diluted $Q$ -- component face--cubic model in two
dimensions has revealed existence of four competing possible types
of critical behavior corresponding to $Q$ -- state Potts model, $2Q$
-- state Potts model, Ising model and special "cubic" fixed point
\cite{rie81, nie83}. It was also found that at $Q<Q_c=2$ the
transitions are continuous and critical behavior of the discrete
face--cubic model belongs to the O(N)-model universality class. For
$Q=2$ the Ashkin--Teller--like behavior occurs and for $Q>2$
transitions are of first order. There are also early results
obtained within the mean--field (MF) theory \cite{kim75} and using
the Bethte--Peirels (BP) approximations and high--temperature series
\cite{kim75(2)} for the case when only dipolar pair interactions are
included $(K=0)$. In the BP approximation the critical value of spin
component $Q_c$, above which transitions are of the first order was
found to be given by
\begin{eqnarray}
Q_c=1+\frac{2}{3}q \left[ \left(1+\frac{6}{q} \right)^{1/2}-1
\right], \label{1.8}
\end{eqnarray}
where $q$ is the coordination number of the lattice. The limit $ q
\to \infty$ corresponds to the MF solution and gives $Q_c=3$. In
Ref. \cite{kim75(2)} the high--temperature series for $Q$ --
component face--cubic model on three dimensional fcc lattice were
constructed up to 5-th order, from which authors obtained $Q_c=2.35
\pm 0.2$, whereas Eq. (\ref{1.8}) gives us $Q_c \simeq 2.8$
Moreover, although the MF solution of cubic model predicted the
tricritical like behavior \cite{kim75}, the BP approximation showed
that it is not the case \cite{kim75(2)}. Only inclusion of
single--ion--anisotropy terms, quadrupolar pair interactions and
crystal fields may drive the system tricritical \cite{kim76}.

 The paper is organized as follows. In Section II we give the
     introduction to recursive lattice constructions and their role in
     statistical mechanics. Then we discuss the peculiarities of the
     Cayley tree, Bethe lattice and connections between these two
     objects. Section III is devoted to the graph expansions for the
     partition function of the model under consideration defined
     on the arbitrary planar graph. We presented the partition
     function in terms of double power series which in case of the
     Cayley tree are summed up exactly, giving the same expression as
     in case of one--dimensional linear chain. We also discuss
     arguments of Eggarter concerning the mechanisms of phase
     transitions on the Cayley tree and Bethe lattice. In the next
     Section IV we derive the system of recurrent relations for the
     $FC_Q$-model on the Bethe lattice reformulating the
     statistical--mechanical problem in terms of the theory of
     dynamical systems \cite{shu}. We introduce the order
     parameters and give their expressions in terms of recursive
     scheme. In Section V we identify different types of the
     fixed points of the system of recurrent relation with
     different physical phases and present the plots of order
     parameters temperature dependencies. The phase diagrams for
     different values of spin component number $Q$ are also
     presented. Brief summary is given in Section VI.

\section{Recursive methods, Cayley tree and Bethe lattice}
Among the vast variety of statistical mechanics lattice models
 with strong local interactions only very limited amount allows
 exact solutions. These exact solutions are known only for
 low--dimensional systems, more precisely for $d=1$ and $d=2$;
 most of exact solutions for two--dimensional systems are known
 only in the absence of the external field and/or for the special
 choice of model parameters \cite{bax}. Well known conventional
 approximate methods like MF theory and BP approximation in
 general can provide only more or less qualitatively satisfactory
 picture and in some cases they just fail. That is why the
 quest for the alternative approaches, which can provide more
 reliable results for the thermodynamics of lattice models is very
 important.

  \begin{figure}
           \begin{center}
           \includegraphics{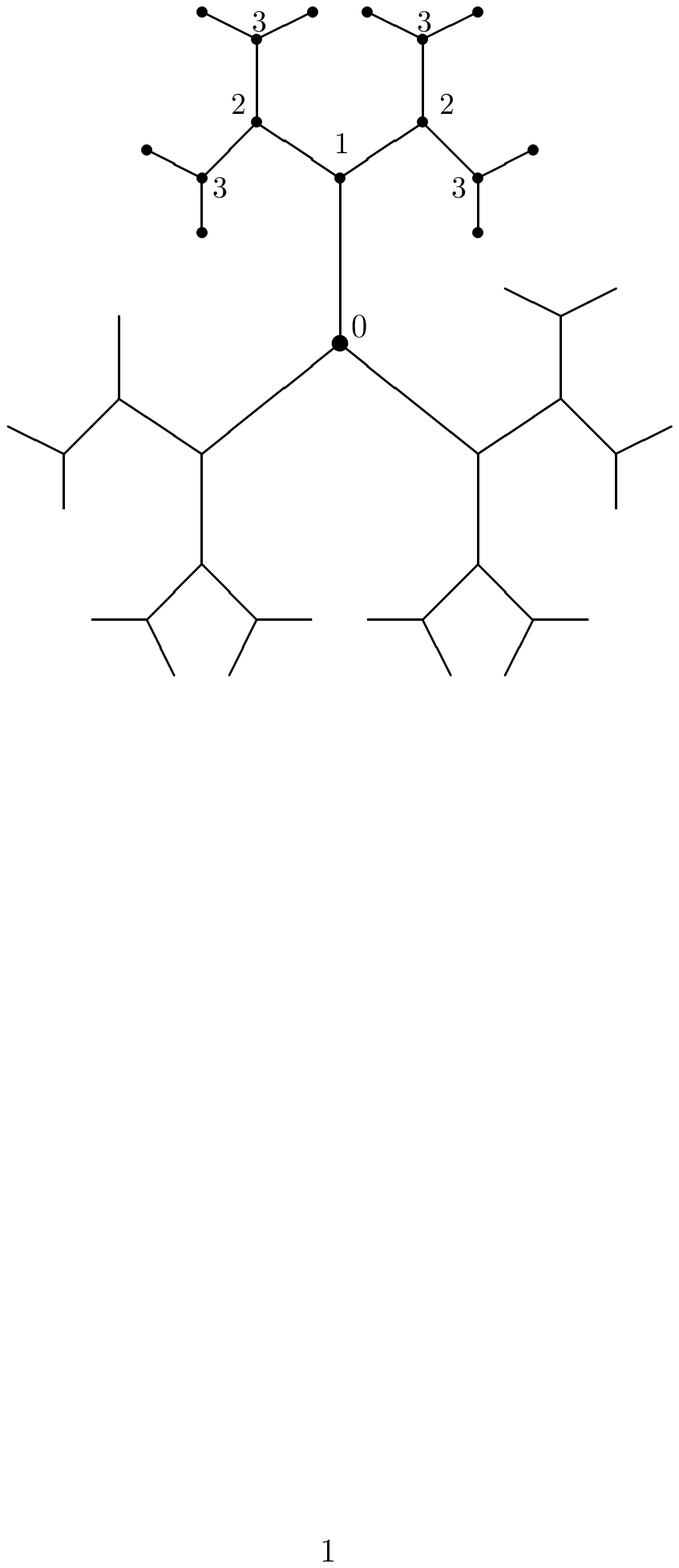}
           \caption{\label{fig1}  The Cayley tree with coordination number $q=3$ and 3 shells.}
           \end{center}
       \end{figure}

  The recursive lattices are of twofold interest. On the one hand,
  the models, defined on a recursive lattices can be considered as
  an independent original area of research in which the powerful
  methods of dynamical systems theory and fractal geometry are
  successfully exploited for determining the thermodynamical
  properties of the statistical mechanics models. On the other
  hand, recursive lattices provide a
  specific kind of approximate treatment of many--particle systems physics.
  In the heart of these approximations not the simplification
  of the character or/and strength of the interaction between the
  system elements, but the modification of the topology of the
  underlying lattice lies. This modification in
  most cases consist in the replacement  of the regular periodic
  "physical" lattice by a recursive one, constructed by the certain algorithm
  and possessing the self--similarity,  maintaining all
  interaction unchanged. For the large class of lattice models
  with commuting variables ("classical models") this approach
  leads
  to the exact solution of the statical mechanics problem in
  terms of the theory of dynamical systems or, more precisely, in terms
  of discrete maps. The solution formally is rather similar to
 that which corresponds to conventional BP approximation but, in
 contrast to the latter, it is the exact solution for the lattice
 endowed with the recursive structure provided the boundary sites
 are properly taken into consideration. It was argued that in some
 cases, particularly in the models where multi--site interactions
 are presented \cite{mon91}-\cite{ara03} the recursive lattice
 approximation gives more reliable results than conventional BP \cite{guj95}.

  The simplest example of recursive lattices is a Cayley tree,
  the self--similar graph, which contains no cycles. In order to
  construct a Cayley tree one should start with a central site
  $O$. At the first step this site should be connected by links
  with $q$ others, which constitute the first shell or the first
  generation of the Cayley tree. The second shell of the Cayley
  tree is constructed by repeating this procedure for all sites of
  the first shell: each of the $q$ sites belonging to the first
  shell connects to the $q-1$ new sites. Thus, the second shell
  contains $N_2=q(q-1)$ sites. Accomplishing this construction for
  n steps one arrives at the recursive connected graph, which
  contains no cycles and is called the Cayley tree with coordination
  number $q$ and $n$ generations (shells) (Fig. \ref{fig1}). As the
  number of sites in $k$--th shell is $q(q-1)^{k-1}$ the total
  number of sites then
  \begin{eqnarray}
N_n=1+\sum_{k=1}^n q \left(q-1 \right)^{k-1}=\frac{q \left(q-1
\right)^n-2}{q-2}. \label{Nn}
  \end{eqnarray}
  Thus, each site of the Cayley tree has coordination number
  equal to $q$ except the sites on the last shell which have only
  one neighbor. The peculiar property of the Caylet tree is the large
  amount of boundary sites which even in the thermodynamical limit
  $n \to \infty$ comprise a finite fraction of the total number of
  sites
  \begin{equation}
  \lim_{n \to \infty}\frac{q\left( q-1 \right)^{n-1}}{N_n}=\lim_{n \to
  \infty}\frac{\left( q-2 \right) q \left( q-1 \right)^{n-1} }{q \left(
  q-1 \right)^n-2}=\frac{q-2}{q-1}. \label{fra}
  \end{equation}
  This feature causes anomalous properties of the statistical
  mechanics systems on the full Cayley tree. For instance, in
  Ising model  there is no zero field magnetization, whereas the
  derivations of free energy with respect to the external field
  exhibit singular behavior \cite{egg74} - \cite{sto03}.

   The Bethe lattice is an object intimately linked to the
   Cayley tree but, in contrast to the latter, it is devoid of the
   complications which arise from the boundary sites. When one
   deals with the Bethe lattice the general structure and topology
   of the Cayley tree are preserved whereas only contributions from
   the bulk sites lying deep inside the Cayley tree are taken into
   consideration. In this regard the Bethe lattice is the interior
   of the Cayley tree \cite{bax, che74}. Dealing with the Bethe
   lattice we suppose the underlying Cayley tree to be large
   enough to achieve thermodynamical limit and consider only the
   sites situated far away from a boundary. One can regard these
   sites as a uniform lattice with coordination number $q$.

    An undeniable advantage of the Bethe lattice is the
    possibility of exact solutions for many types of statistical
    mechanics problems. A large amount of problems of the theory
    of magnetism, macromolecule physics, lattice gauge theory,
    self--organized criticality, dynamical mean--field theory (DMFT) and general questions of
    statistical mechanics like zeroes of partition functions have
    been successfully considered on Bethe or Bethe-like lattices,
    revealing many interesting exact results and deep connections
    between the theory of dynamical systems and statistical
    mechanics \cite{egg74} - \cite{ala98}.

  \section{Face-cubic model on planar graphs}
It is possible to calculate exactly the partition function of the
$FC_Q$-model on Cayley tree in case of the absence of external
field. Let us consider the partition function of the model without
external field on arbitrary planar graph $G$
\begin{eqnarray}
{\mathcal{Z}}_G=\sum_{\{ \sigma \}} \sum_{\{ \alpha \}}
\prod_{\langle i,j\rangle \in G}\exp \left\{ (J\sigma_i \sigma_j +
K)\delta_{\alpha_i, \alpha_j} \right\}. \label{2.2}
\end{eqnarray}
As in the case of Potts model \cite{wu} one can represent it in the
following way:
\begin{eqnarray}
{\mathcal{Z}}_G=\sum_{\{ \sigma \}} \sum_{\{ \alpha \}}
\prod_{\langle i,j\rangle \in G}(1+U_{\sigma_i
\sigma_j}\delta_{\alpha_i, \alpha_j}), \label{2.3}
\end{eqnarray}
where $U_{\sigma_i \sigma_j}=e^{J\sigma_i \sigma_j+K}-1$. Then
summing out over the all $\alpha_i$ we obtain that
\begin{eqnarray}
{\mathcal{Z}}_G=\sum_{\{ \sigma \}}\sum_{G^{\prime}\subseteq G}
Q^{R_0(G^{\prime})}\prod_{\langle i,j\rangle \in
G^{\prime}}U_{\sigma_i \sigma_j}, \label{2.4}
\end{eqnarray}
where the second sum is going over all spanning graphs $G^{\prime}$
of the underlying planar graph $G$ and $R_0(G^{\prime})$ is the 0-th
Betti's number of $G^{\prime}$ which coincides with the number of
connected components of $G^{\prime}$ and the product is over all
links $\langle i,j\rangle$ belonging to $G^{\prime}$. Using the
identity $\sigma_i \sigma_j = 2 \delta_{\sigma_i \sigma_j}-1$ one
can represent Eq. (\ref{2.4}) in terms of double power series
associated with the underlying lattice provided latter is a planar
graph:
\begin{eqnarray}\nonumber
{\mathcal{Z}}_G&=&\sum_{G^{\prime}\subseteq
G}Q^{R_0(G^{\prime})}u^{e(G^{\prime})}\sum_{\{ \sigma
\}}\prod_{\langle i,j\rangle \in G^{\prime}}(1+v\delta_{\sigma_i
\sigma_j}), \\ \label{2.5} u&=&e^{K-J}-1, \nonumber \\
v&=&\frac{e^{2J}-1}{1-e^{J-K}},
\end{eqnarray}
where $e(G^{\prime})$ is the edges number in the graph $G^{\prime}$
and all $\sigma_i$ assume values 0 or 1. So, for each spanning
subgraph $G^{\prime}$ of planar graph $G$ we obtain a partition
function of 2-state Potts model:
\begin{eqnarray}\nonumber
{\mathcal{Z}}_G&=&\sum_{G^{\prime}\subseteq
G}Q^{R_0(G^{\prime})}u^{e(G^{\prime})}{\mathcal{Z}}_{G^{\prime}}^{Potts}(v)
\\ &=&\sum_{G^{\prime}\subseteq
G}Q^{R_0(G^{\prime})}u^{e(G^{\prime})}\sum_{G^{\prime
\prime}\subseteq G^{\prime}}2^{R_0(G^{\prime \prime})}v^{e(G^{\prime
\prime})}, \label{2.6}
\end{eqnarray}
where the second sum is going over all subgraph of $G^{\prime}$
 Thus, we have succeeded in representing the partition function of the $FC_Q$ -
 model given by Hamiltonian (\ref{1.6}) in terms of double power
 series associated with the underlying lattice provided the latter
 is a planar graph.
 As in case of ordinary spin Potts model one can make
 $1/Q$-expansions for $FC_Q$-model with the aid of Eq.
 (\ref{2.6}). For instance, for the two--dimensional square lattice one
 obtains
 \begin{eqnarray}
{\mathcal{Z}}&=&(2Q)^N
\{1+2N\frac{u}{Q}(\frac{v}{2}+1)+N(2N-1)(\frac{u}{Q})^2(\frac{v}{2}+1)^2+C_{2N}^3
(\frac{u}{Q})^3(\frac{v}{2}+1)^3 \\ \nonumber
  &+& (C_{2N}^4 - N)(\frac{u}{Q})^4(\frac{v}{2}+1)^4+N
  \frac{u^4}{Q^3}((\frac{v}{2}+1)^4 + (\frac{v}{2})^4)+(C_{2N}^5 - N(2N-4))(\frac{u}{Q})^5 (\frac{v}{2}+1)^5 \\ \nonumber
  &+& N(2N-4)\frac{u^5}{Q^4}((\frac{v}{2}+1)^5+(\frac{v}{2})^4+(\frac{v}{2})^5) \\ \nonumber
  &+&
  2N\frac{u^7}{Q^5}((\frac{v}{2}+1)^7+2((\frac{v}{2})^4+(\frac{v}{2})^5)+3(\frac{v}{2})^6)+ \ldots
  \}.
 \end{eqnarray}

 Now we can use the well known relations
 between the topological invariants of graphs \cite{graph} to
 rewrite Eq. (\ref{2.6}) in another way suitable for our further
 purposes. Applying the Euler theorem to planar graph $G$ we get
 \begin{eqnarray}
 R_0(G)-R_1(G)=n-e(G), \label{2.7}
 \end{eqnarray}
 where the first Betti's number $R_1(G)$ coincides with the number
 of independent cycles of $G$ and $n$ is the number of sites.
 Therefore
 \begin{equation}
{\mathcal{Z}}_G=(2Q)^n\sum_{G^{\prime}\subseteq
G}Q^{R_1(G^{\prime})}(\frac{u}{Q})^{e(G^{\prime})}\sum_{G^{\prime
\prime}\subseteq G^{\prime}}2^{R_1(G^{\prime
\prime})}(\frac{v}{2})^{e(G^{\prime \prime})}. \label{2.8}
 \end{equation}
 \subsection{The case of graphs without cycles: trees and forests}
 Partition function for $FC_Q$-model in the form of Eq.
 (\ref{2.8}) can easily be calculated in case of the so-called
 forests, the planar graphs, which contain no cycles. Connected
 part of forest is called tree. Let us suppose that we have a tree
 $T$ with $n$ sites and $L_n$ edges. Then Eq.(\ref{2.8}) takes the
 form
 \begin{eqnarray}\nonumber
{\mathcal{Z}}_T&=&(2Q)^n\sum_{G^{\prime}\subseteq
T}(\frac{u}{Q})^{e(G^{\prime})}\sum_{G^{\prime \prime}\subseteq
G^{\prime}}(\frac{v}{2})^{e(G^{\prime \prime})} \\
&=&(2Q)^n\sum_{k=0}^{L_n}C^k_{L_n}(\frac{u}{Q})^k\sum_{l=0}^kC^l_k(\frac{v}{2})^l,
\label{2.9}
 \end{eqnarray}
 where $C^l_k=\frac{k!}{l! (k-l)!}$ are the binomial coefficients
 and the sum is going over all subgraphs with given volume $k$.  With the aid of Newton's binom we obtain
 \begin{eqnarray}
{\mathcal{Z}}_T=(2Q)^n \left( \frac{u}{Q} \left( \frac{v}{2}+1
\right)+1 \right)^{L_n}. \label{2.10}
 \end{eqnarray}
If $T$ is the Cayley tree then $L_n=n-1$ and we will have
\begin{eqnarray}
{\mathcal{Z}}_{Cayley}=2Q \left(u \left( v+2 \right)+2Q
\right)^{n-1}. \label{2.11}
\end{eqnarray}
Thus, the free energy per site for $FC_Q$-model on Cayley tree is

\begin{eqnarray}\nonumber
f=-k_BT \lim_{n \to \infty}\frac{\log{\mathcal{Z}}_{Cayley}}{n}=
-k_BT\log\left(u \left( v+2 \right)+2Q \right)\\
=-k_BT\log2-k_BT\log\left(e^K\cosh J+Q-1 \right).
\end{eqnarray}

 It is noteworthy that this result is in full agreement
with that obtained by Aharony for one--dimensional model by
transfer--matrix technique \cite{aha77}. Indeed, a one--dimensional
chain can be regarded as a "tree" in the sense mentioned above.
Thus, the free energy of the $FC_Q$ model on a Cayley tree in a
thermodynamic limit is continuous for all $T$. Formally it coincides
with that for one--dimensional systems where the continuity is
really the case. But, as is known, the BP approximations for the
lattice model with an arbitrary number of component solves exactly
the problem on a Cayley tree \cite{whe70}. Therefore the model on a
Cayley tree must possess a finite critical temperature as predicted
by BP approximation \cite{kim75(2)}. The origin of this peculiar
properties of the Cayley tree was revealed by Eggarter \cite{egg74}
on the example of Ising model. He argued that the equivalence of all
sites in the thermodynamic limit, which is one of the key points of
BP approximation, breaks down for the sites of the Cayley tree which
are situated close to the surface. As appears from Eq. (\ref{fra})
these sites comprise rather large fraction of the sites of the
Cayley tree and, thus, they determine the behavior of all
thermodynamic quantities to a great extent.

 However, even though no phase transitions in the thermodynamic
 limit occur at the value of critical temperature $T_c$ predicted
 by the BP method the value of the order parameter (magnetization
 etc.) in the interior of the lattice, i. e. for the region far
 from the surface, which is generally called "Bethe lattice", will undergo jumps from zero to its BP value
 when the temperature is below $T_c$. In order to
 illustrate it we will use the technique of dynamical system
 theory, which becomes a powerful tool for investigating various
 physical problems on recursive lattices.

 \section{Recursive method for Face Cubic model on Bethe lattice.}
    Many statistical systems defined on the recursive lattices are famous
 for the possibility of exact solution in terms of dynamical
 system theory. In the heart of these exact solutions the
 self-similarity of recursive lattices lies. For instance, if we cut
 apart the Cayley tree at the central site it will give q identical
 branches each of which is the same Cayley tree with the number of
 generation decreased by 1. Using this fact one can establish
 the connection between the partition function of a model defined
 on the Cayley tree containing n shells with the partition
 function of the same model defined on the Cayley tree containing
 n-1 shells. Therefore, the thermodynamical problem is
 reformulated in terms of discrete maps, given by recurrent
 relations. Let us introduce a symmetry breaking field into
 Hamiltonian (\ref{1.6}). The field can be considered as a uniform
 magnetic field pointing along the first coordinate axis
 ${\mathbf{H}}=\left(H,0,...0 \right)$. The corresponding
 interaction Hamiltonian is of the conventional Zeeman type:
 \begin{eqnarray}
 {\mathcal{H}}_Z=-{\mathbf{H}}\sum_i {\mathbf{S}}_i, \label{zee}
 \end{eqnarray}
 which leads to the following form of the $FC_Q$-model Hamiltonian
 in the field:
 \begin{equation}
 - \beta{\mathcal{H}}_{FC}=J\sum_{\langle i,j\rangle}\sigma_i \sigma_j
 \delta_{\alpha_i, \alpha_j}+K\sum_{\langle i,j\rangle }\delta_{\alpha_i,
 \alpha_j}+h \sum_i \sigma_i \delta_{\alpha_i , 1}, \label{3.0}
 \end{equation}
 where $h= \beta H$. Hereafter we pass from the Cayley tree to the Bethe
 lattice having in mind one of the ways mentioned in Section II.
 Thus, we should no longer care about boundary sites and boundary
 conditions.
 According to that one can represent the partition function
 of the  $FC_Q$-model on the Bethe lattice in the following form:
 \begin{eqnarray}
 Z=\sum_{(\sigma_0,\alpha_0)}e^{h\sigma_0\delta_{\alpha_0,1}}[g_n(\sigma_0,
 \alpha_0)]^q, \label{3.1}
 \end{eqnarray}
 where $\sigma_0$ and $\alpha_0$ are the variables of the spin in
 the central site and $g_n(\sigma_0,
 \alpha_0)$ refers to a partition function of individual branch:

 \begin{eqnarray}
 g_n(\sigma_0, \alpha_0)=\sum_{\sigma\neq\sigma_0}\sum_{\alpha\neq\alpha_0}
 \exp \left(\left(J\sigma_0\sigma_1+K \right)\delta_{\alpha_0, \alpha_1}+
 \sum_{\langle i,j\rangle}\left(J\sigma_i\sigma_j+K \right)\delta_{\alpha_i, \alpha_j}+
 \sum_i \sigma_i \delta_{\alpha_i, 1} \right). \label{3.2}
 \end{eqnarray}

 Each branch, in its turn, can be cut at the site which was
 previously connected to the central one. This will give us q-1
 identical branches being the Bethe lattices with n-1 generations.
 Thus, the connection between $g_n$ and $g_{n-1}$ is

 \begin{eqnarray}
g_n(\sigma_0,
 \alpha_0)=\sum_{(\sigma_1, \alpha_1)}\exp\left(\left(J\sigma_0\sigma_1+K \right)\delta_{\alpha_0,
 \alpha_1}+ h\sigma_1\delta_{\alpha_1,1}\right)[g_{n-1}(\sigma_1,
 \alpha_1)]^{q-1}, \label{3.3}
 \end{eqnarray}

 Here we obtain the system of $2Q$ recursion relations but, in
 fact, only three of them are independent as all $g_n(\sigma,
 \alpha)$ for $\alpha\neq1$ are identical. So, from (\ref{3.3}) we
 have

 \begin{eqnarray}
 g_n(+,
 1)&=&e^{J+K+h}[g_{n-1}(+,1)]^{q-1}+e^{-J+K-h}[g_{n-1}(-,1)]^{q-1}+2(Q-1)[g_{n-1}(\pm,*)]^{q-1},
 \\ \nonumber
g_n(-,
 1)&=&e^{-J+K+h}[g_{n-1}(+,1)]^{q-1}+e^{J+K-h}[g_{n-1}(-,1)]^{q-1}+2(Q-1)[g_{n-1}(\pm,*)]^{q-1},
 \\ \nonumber
 g_n(\pm,
 *)&=&e^{h}[g_{n-1}(+,1)]^{q-1}+e^{-h}[g_{n-1}(-,1)]^{q-1}+(e^{J+K}+e^{-J+K}+2(Q-2))[g_{n-1}(\pm,*)]^{q-1},
 \label{ggg}
 \end{eqnarray}

where $g_n(\pm,*)$ stand for any 2Q-2 partition functions
corresponding to the individual branch with $\vec{S}_1$ whose
direction is non collinear with the first coordinate axes.
Introducing the variables
\begin{eqnarray}
x_n=g_n(+,1)/g_n(\pm,*),
\nonumber \\
y_n=g_n(-, 1)/g_n(\pm,*), \label{xyg}
\end{eqnarray}
we obtain the system of two recurrent relations
\begin{eqnarray}\nonumber
x_n=f_1(x_{n-1}, y_{n-1}), \\  y_n=f_2(x_{n-1}, y_{n-1}),
\label{rr1}
\end{eqnarray}

with

\begin{eqnarray}
\label{rrpuk}
f_1(x,y)=\frac{P_1(x,y)}{R(x,y)}=\frac{a\mu x^{q-1}+b
\mu^{-1}y^{q-1}+2(Q-1)}{\mu x^{q-1}+\mu^{-1}y^{q-1}+a+b+2(Q-2)},
\\ \nonumber f_2(x,y)=\frac{P_2(x,y)}{R(x,y)}=\frac{b\mu x^{q-1}+a
\mu^{-1}y^{q-1}+2(Q-1)}{\mu x^{q-1}+\mu^{-1}y^{q-1}+a+b+2(Q-2)},
\end{eqnarray}

where the following notations are adopted
\begin{eqnarray}
a&=&\exp(J+K),\\ \nonumber b&=&\exp(-J+K), \\ \nonumber \mu
&=&\exp(h). \label{33}
\end{eqnarray}
In this approach statistical averages of all physical quantities can
be expressed in terms of $x$ and $y$ variables defined by Eq.
(\ref{rr1}). For instance, when the total number of spins is $n$ the
magnetization along the first coordination axis which is the thermal
average of the following form:
\begin{eqnarray}
m=\frac{1}{n}\sum_{i=1}^n\langle
S_i^{(1)}\rangle=\frac{1}{n}\sum_{i=1}^n\langle \sigma_i
\delta(\alpha_i, 1 )\rangle, \label{3.10}
\end{eqnarray}
in terms of $x$ and $y$ are expressed as
\begin{eqnarray}
m=\frac{\mu x^q - \mu^{-1} y^q}{\mu x^q +\mu^{-1}y^q+2(Q-1)},
\label{m}
\end{eqnarray}
provided all sites of the lattice are equivalent. It is easy to see
that this quantity plays the role of an order parameter of Ising
universality class. Another order parameter belonging to the
$Q$--state Potts universality class is
\begin{eqnarray}
p=\frac{1}{n}\sum_{i=1}^n\langle \delta(\alpha_i, 1 )\rangle,
\label{p} \label{3.12}
\end{eqnarray}
which also can be regarded as a "quadrupolar" moment $\langle {
S_i^{(1)}}^2 \rangle$. Thus
\begin{eqnarray}
p=\frac{\mu x^q + \mu^{-1} y^q}{\mu x^q
+\mu^{-1}y^q+2(Q-1)}.\label{3.13}
\end{eqnarray}

  The order parameters $m$ and $p$ define the three possible
  phases of the $FC_Q$ model in case of the ferromagnetic
  couplings, i.e. $J>0, K>0$, provided external magnetic field is vanished:\\
   \\
   (a) disordered (paramagnetic) phase:  $m=0$, $p=1/Q$  \\
   \\
   (b) ferromagneticaly ordered phase: $m\neq0$ , $p\neq1/Q$ \\
   \\
   (c) partially ordered (quadrupolar) phase: $m=0$ , $p\neq1/Q$

 \section{Investigations of phase transitions in terms of dynamical systems theory}
 In order to obtain physical results one must implement an
 iterative procedure for the RR
 (\ref{rrpuk}). Namely, starting from the random initial
 conditions $(x_0, y_0)$ one uses the simple iterations scheme and
 examines the behavior of physical quantities after a large number
 of iterations \cite{van81}. In a simplest case  the iterative
 sequence $\{x_n, y_n\}$ converges to a fixed point $(x^*, y^*)$,
 which is defined by
 \begin{eqnarray}
 \left\{ \begin{array}{rl}
 & x^*=f_1(x^*,y^*) \\
 & y^*=f_2(x^*,y^*)
 \end{array} \right.
 \end{eqnarray}
 This situation is inherent to the ferromagnetic case when both $J$
 and $K$ are positive.
 As can easily be seen from Eqs. (\ref{3.10}) - (\ref{3.13}), when the
 magnetization and quadrupolar moment of the system take values
 $m$ and $p$ respectively, then the corresponding fixed point of
 the RR (\ref{rrpuk}) is the following,
 provided $h=0$:
 \begin{eqnarray}
 \left\{ \begin{array}{rl}
 & x^*=\left( \frac{(Q-1)(p+m)}{1-p} \right)^{1/q}\\
 & y^*=\left( \frac{(Q-1)(p-m)}{1-p} \right)^{1/q}
 \end{array} \right.
 \end{eqnarray}
 Thus, according to the properties of different phases of the
 ferromagnetic
 $FC_Q$ model one can establish the connection between them and the
 classification of the possible types of fixed points:\\
   \\
   (a) disordered (paramagnetic) phase:  $x^*=y^*=1$  \\
   \\
   (b) ferromagneticaly ordered phase: $x^* \neq y^* \neq 1$ \\
   \\
   (c) partially ordered (quadrupolar) phase: $x^*=y^* \neq 1$\\
   \\
 The condition
 $x^*=y^*$ leads to $g_n(+,1)=g_n(-,1)$, which means that the probability
 of spin "up" is equal to the probability of spin "down" which is really the
  case in paramagnetic phase.
Applying the simple iterative scheme to Eqs. (\ref{rr1}), (\ref{m})
and (\ref{3.13}) one can obtain the plots of magnetization processes
($m$ vs. $H$ at fixed values of $T$) as well as the zero field
magnetization and quadrupolar moment of the system.
\begin{figure}
\begin{center}
\begin{tabular}{cc}
{\small (a)}&{\small (b)}\\
\includegraphics{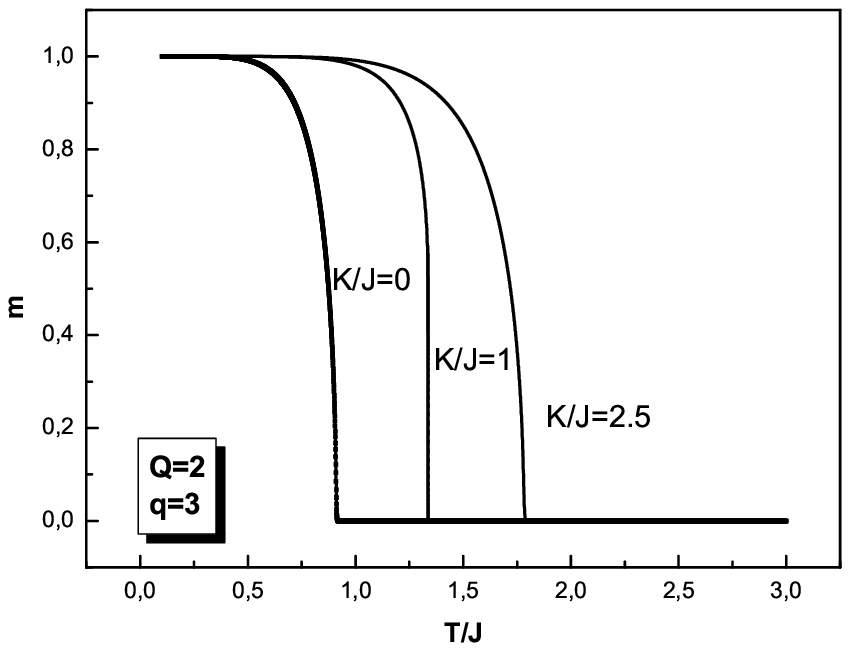}&\includegraphics{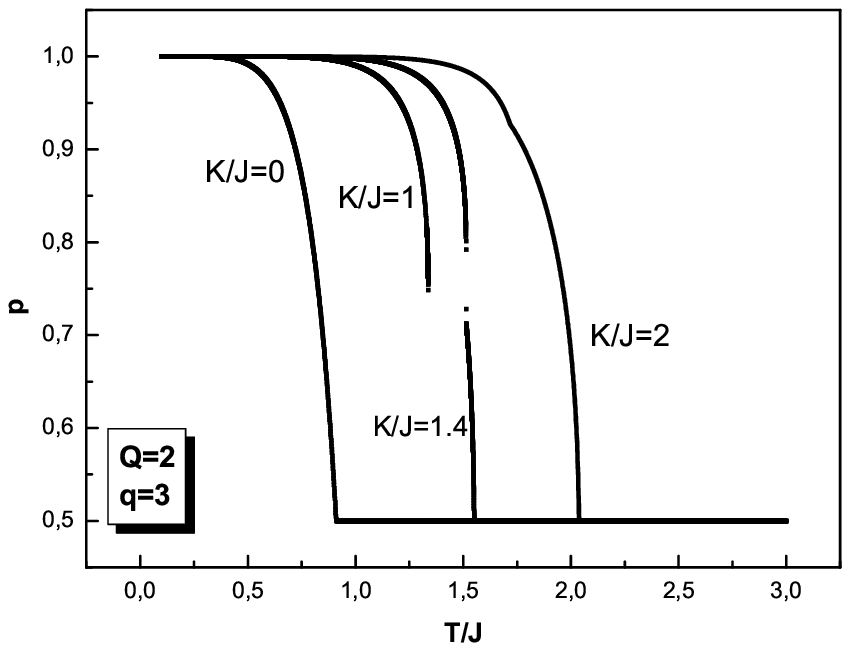}\\
\end{tabular}
\caption {\small {The temperature behavior of the order parameters
for $Q=2$ and $q=3$. The transitions from disordered to
ferromagnetic phase is of second order for all range of $K$ .(a)
Zero field magnetization for $K/J=0; 1$ and 2.5; (b) Quadrupolar
moment in zero field for $K/J=0;1;1.4$ and 2. For $K/J=1$ one can
see the first order transition between disordered and ferromagnetic
phases, whereas at $K/J=1.4$ one can see two subsequent phase
transitions, the continuous transition from disordered to
quadrupolar phase and very close to that the first order transition
to the ferromagnetic phase. With the further increase of $K/J$ the
second transition becomes of the second order.}}
 \label{fig2}
 \end{center}
\end{figure}

 In Fig. \ref{fig2} one can see the temperature dependencies
 of the order parameters of the $FC_Q$-model on Bethe lattice with
 coordination number $q=3$ and $Q=2$. As is obviously seen from
 Fig. \ref{fig2}(a) in this case the zero field magnetization is
 always continuous, whereas the behavior of $p$ is quite different.
 Depending on the value of ratio $K/J$ the $p(T)$ curve could be
 continuous or can include one first order transition
 point. In Fig. \ref{fig2}(b) one can see that at small values of $K/J$ only the one phase
 transition occurs from disordered phase to ferromagnetic phase omitting the partially ordered phase.
This transition is of the second order until $K/J$ reaches some
intermediate value above which one can see a discontinuity in the
order parameter $p$ at critical temperature corresponding to the
transition between disordered and ferromagnetic phases.
 At large values of $K/J$ the system undergoes successively two
 phase transitions at temperatures $T_q(K)$ and $T_f(K)$, the
 larger one corresponds to the transition between disordered and
 quandrupolar phases, the lower one -- to the transitions from
 quandrupolar to ferromagnetic. Within this region of the values
 of $K$ one can see the second order transition from disordered
 phase into "quadrupolar" phase and the further first order transition
 to the ferromagnetic phase. However, beginning with some value of
 $K$ the transition between quadrupolar and ferromagnetic phases
 becomes continuous.
  \begin{figure}
 \begin{center}
\includegraphics{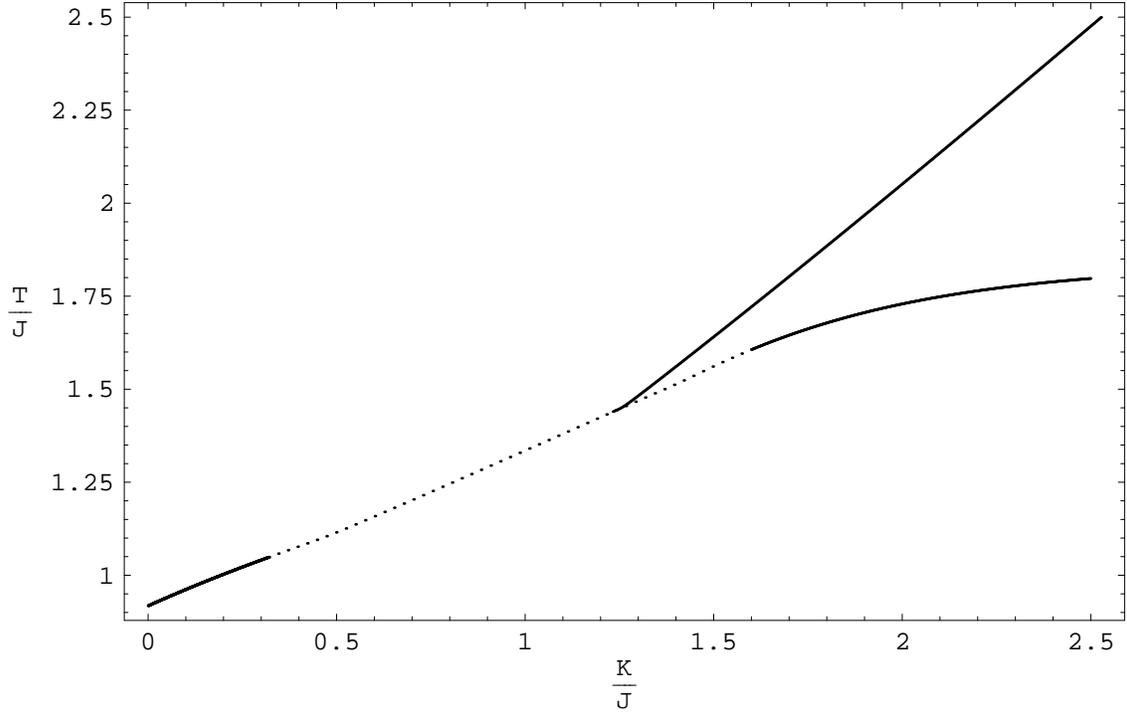}
   \end{center}
  \caption{ The phase diagram of the model for $Q=2$ and $q=3$. Solid
line corresponds to the second order transitions, dotted - to the
first order transitions. For small values of the ratio $K/J$ one can
see only single phase transition line between disordered
hight--temperature phase and completely ordered ferromagnetic phase,
whereas for $K/J\geq 1.233$ there is the partially ordered
"quadrupolar" phase between them. The tricritical points are
$K/J\approx 0.32;1.233$ and 1.6. \label{fig.3} }
\end{figure}

  In order to complete the picture of the phase structure of the model
  under consideration at $Q=2$ we plot a phase diagram by
  separating the regions of the fixed points of different kinds of the
  RR
  from Eq. (\ref{rrpuk}) in the $(K/J, T/J)$-plane. One can see in
   Fig. \ref{fig.3} that in case of $Q=2$, $q=3$ at $K=0$ the
  phase transition from disordered to ferromagnetic phase is of the
  second order. This feature maintains up to $K/J\approx 0.32$,
  where one can notice the tricritical point separating the region
  of continuous transitions from the region of the first order
  transitions. The line of the second order phase transitions between
  disordered and quadrupolar phases merges the line of
  transitions between disordered and ferromagnetic phase at $K/J\approx
  1.233$, the latter, in its turn, again becomes of the second order at
  $K/J\approx1.6$ and for large values of $K/J$ it becomes parallel
  to the $K/J$ axis which means that the transition temperature
  between quadrupolar and ferromagnetic phases is unaffected by
  the value of $K$ beginning with $K/J\approx4$.
\begin{figure}
\begin{center}
\begin{tabular}{cc}
{\small (a)}&{\small (b)}\\
\includegraphics{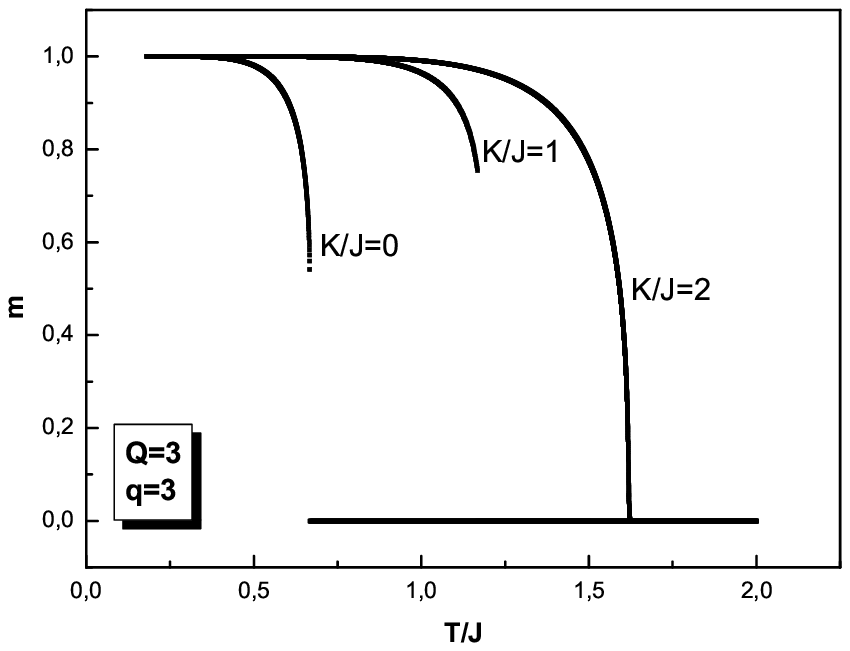}&\includegraphics{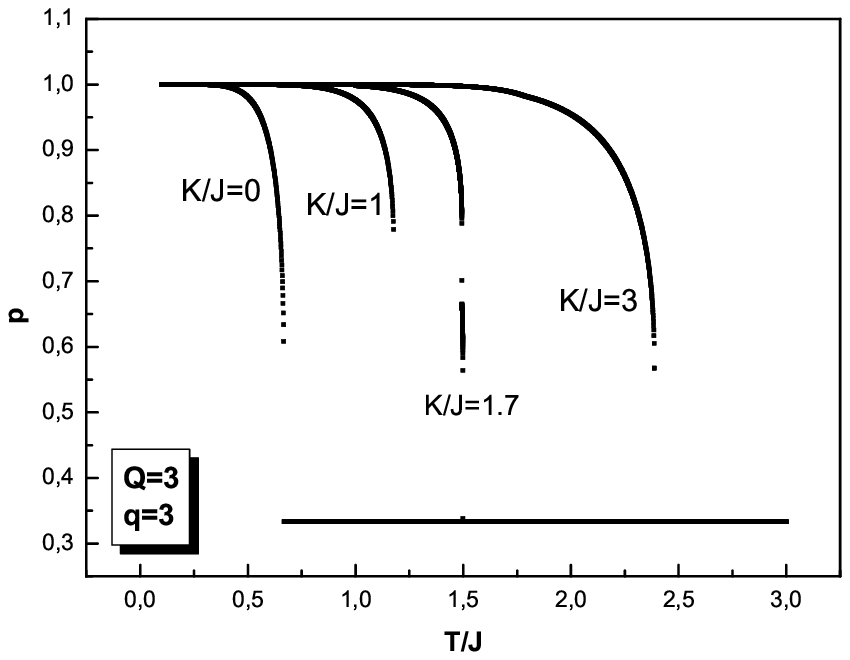}\\
\end{tabular}
\caption {\small {The temperature behavior of the order parameters
for $Q=3$ and $q=3$. Here magnetization of the system undergoes a
jump for small values of $K/J$ and becomes continuous beginning with
$K/J\approx 1.84$. Another order parameter always remains
discontinuous at the transition points. In the plots $p(T)$ for
$K/J=1.7$ one can see that the transitions between "quadrupolar" and
ferromagnetic phases is of the first order, whereas with the further
increase of $K/J$ in becomes continuous. }}
 \label{fig.4}
 \end{center}
\end{figure}

\begin{figure}
\begin{center}
\centerline{\includegraphics{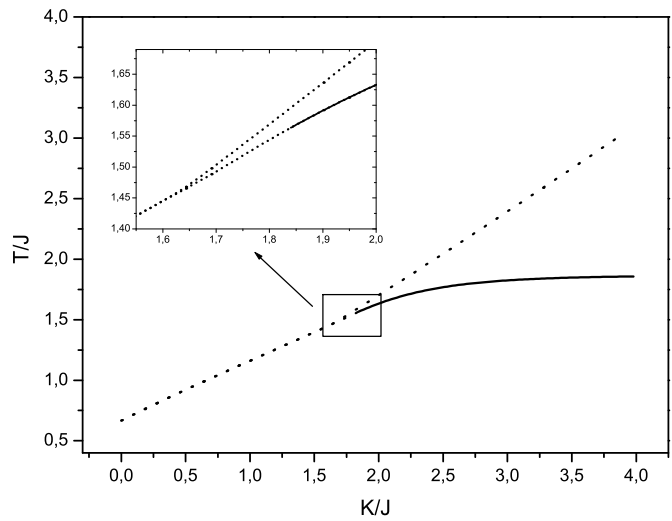}} \caption{The phase diagram
of the model for $Q=3$ and $q=3$. Solid line corresponds to the
second order transitions, dotted - to the first order transitions.
The inset shows the merge point of the three first order lines
(triple point): disordered--ferromagnetic, disordered--"quadrupolar"
and "quadrupolar"--ferromagnetic. In this point at $K/J\approx 1.67$
these three phase are in the equilibrium . The tricritical point on
the line between "quadrupolar" and ferromagnetic phases is situated
rather close to the triple point. The distance between them tends to
zero with the increase of $Q$.} \label{fig.5}
\end{center}
\end{figure}

   The situation is quite different for $Q>2$. In Fig.
   \ref{fig.4} the plots of zero field magnetization and
   quadrupolar moment for $Q=3$ and $q=3$ are presented. In this
   case both $m$ and $p$ are discontinuous at the critical
   temperatures. Moreover, this feature is preserved for $p$ for
   all values of $K/J$, whereas $m$ becomes continuous after
   $K/J\approx1.84$. the corresponding phase diagram is presented
   in Fig. \ref{fig.5}. Here one can see a triple point at
   $K/J\approx1.67$ where the merging of the three lines of the
   first order transitions takes place. There is also a
   tricritical point at $K/J\approx1.84$ which separates the
   region of the first and the second order phase transitions on the
   line between quadrupolar and ferromagnetic phases. The line, as
   in case of $Q=2$, goes to plateau beginning with $K/J \approx 5$,
   so the transition temperature again does not depend on $K$ for
   $K \geq 5J$.
     In general, for $Q>3$ the pase diagrams of the model under
     consideration have the same topology as in case of $Q=3$. The
     main feature is the decrease of the distance between triple
     and tricritical points with the increase of the $Q$. So, for $Q=30$
     one can obtain that this distance is of $10^{-2}$ order with $K/J$.
     Apparently, they have never shrunk to a single point at
     finite values of $Q$, but it could be the case when
     $Q\rightarrow\infty$.
      The value of critical spin component number $Q_c$ at which
      the phase transitions become of the first order at $K=0$ in
      the Face-cubic model on Bethe lattice is exactly equal to
      3, which coincides with the mean--field results at
      $q\rightarrow\infty$, but, in contrast to that, on Bethe
      lattice this value is unaffected by the value of
      coordination number $q$. Obviously, this is the direct
      consequence of the fact, that the Bethe--Pierels
      approximation become exact on the Bethe lattice.
      \section{Summary}
      We have considered very general classical spin model with
      cubic symmetry on the Bethe lattice. First of all, for the
      underlying Cayley tree, which is the planar graph, we have
      presented the partition function of the model as a double
      power series in terms of graph expansions for arbitrary
      planar graphs. This double graph expansion representing the known
      formula for linear chain was evaluated
      exactly in case of Cayley tree. The expression for free energy per
      spin in the thermodynamic limit in this case was found to be
      continuous in temperature. This is in full agreement with
      general statement about mechanisms of phase transitions on
      recursive lattices. In the thermodynamic limit the influence of the huge amount of boundary
      sites, which in contrast to all other sites have only one
      neighbor, precludes the system from undergoing any phase
      transitions in the sense usually understood in the thermodynamics of
      "normal" systems. Then, getting rid of all boundary sites by
      formal putting of the coordination number of all sites equal to
      $q$, thus, passing to the completely homogeneous connected recursive graph,
       the Bethe lattice, we have applied the methods of the
       dynamical systems theory or, more precisely, the theory of
       discrete mappings. In this technique one exploits the
       self-similarity of the Bethe lattice and establishes a
       connection between the thermodynamic quantities defined for
       the lattices with different number of sites. We have used
       the method in which this connection is given by the
       RR for some formal variables $x$ and $y$ which have no
       direct physical meaning but can be used to determine any
       thermodynamic function of the system. The key point of
       this approach is the iterative procedure for the system of
       RR $x_n=f_1\left(x_{n-1}, y_{n-1} \right)$, $y_n=f_2\left(x_{n-1}, y_{n-1}
       \right)$, where $n$ can be regarded as the number of
       generations into the Bethe lattice. Thus, doing the
       iterations one approaches the thermodynamic limit $n \to
       \infty$. However, actually the iterative procedure is
       terminated for some finite number of iterations by the
       fixed point (which is the case for ferromagnetic
       couplings)or by s-cycles, which takes place in the
       antiferromagnetic case. There is also more complicated case
       of the chaotic behavior of the iterative sequence but it is
       out of the topic of the article. We have identified the
       different thermodynamic phases of the system in
       ferromagnetic case ($J>0$, $K>0$)
       (disordered, partially ordered and completely ordered) with
       different types of the fixed points of RR. According to
       that we have obtained the phase diagrams of the model which
       are found to be different for $Q\leq 2$ and $Q> 2 $. The
       former case contains three tricritical points whereas the
       latter just one tricritical and one triple points.
 \section{Acknowledgments}
The authors are grateful to N. S. Izmailian for useful discussions.
This work was partly supported by grants ISTC A-655 and ISTC A-820.
L. N. Ananikyan thanks NFSAT (GRSP16/06) for  partly support.


\begin{thebibliography}{99}

\bibitem{aha74} A. Ahahrony and A. D. Bruce, Phys. Rev. Lett.
\textbf{33}, 427 (1974).

\bibitem{bru75} A. D. Bruce and A. Aharony, Phys. Rev.
B \textbf{11}, 478 (1975).

\bibitem{aha76}A. Aharony, in \textit{Phase Transition and Critical
Phenomena}, ed. by C. Domb and J. Lebowitz (Academic Pess, New York,
1976), vol. 6, p. 357.

\bibitem{pel02} A. Pelissetto and E. Vicari, Phys. Rept.
\textbf{368}, 549 (2002).

\bibitem{cal02} P. Calabrese and A. Celi, Phys. Rev. B \textbf{66}, 184410
(2002).

\bibitem{cal03} P. Calabrese, A. Pelissetto, and E. Vicari, Phys. Rev. B \textbf{67}, 054505
(2003).

\bibitem{cal04} P. Calabrese, E. V. Orlov, D. V. Pakhnin, and A. I. Sokolov,
Phys. Rev. B \textbf{70}, 094425 (2004).

\bibitem{bar89} N. C. Bartelt, T. L. Einstein, and L. T. Wille,
Phys. Rev. B \textbf{40}, 10759 (1989); T. Aukrust, M. A. Novotny,
P. A. Rikvold, and D. P. Landau, Phys. Rev. B \textbf{41}, 8772
(1990).

\bibitem{cho93} T. Chou and D. R. Nelson, Phys. Rev. E \textbf{48},
4611 (1993).

\bibitem{bek00} S. Bekhechi, A. Benyoussef, and N. Moussa, Phys.
Rev. B \textbf{61}, 3362 (2000).

\bibitem{kle97} H. Kleinert, S. Thoms, and V. Schulte-Frohlinde,
Phys. Rev. B \textbf{56}, 14428 (1997).

\bibitem{sha97} B. N. Shalaev, S. A. Antonenko, and A. I. Sokolov,
Phys. Lett. A \textbf{230}, 105 (1997).

\bibitem{pak00} D. V. Pakhnin and A. I. Sokolov, Phys. Rev.
B \textbf{61}, 15130 (2000).

\bibitem{pak01} D. V. Pakhnin and A. I. Sokolov, Phys. Rev.
B \textbf{64}, 094407 (2001).

\bibitem{cal02a}  P. Calabrese, A. Pelissetto and E. Vicary, Acta Phys. Slov.
\textbf{52} , 311 (2002).

\bibitem{usa04} Z. Usatenko and J. Spa\l ek, J. Phys. A
\textbf{37}, 7113 (2004).

\bibitem{kac06} H. Kachkachi and E. Bonet, Phys. Rev. B \textbf{
73}, 224402 (2006).

\bibitem{kim75} D. Kim, P. M. Levy and L. F. Uffer, Phys. Rev.
B \textbf{12}, 989 (1975).

\bibitem{kim75(2)} D. Kim and P. M. Levy, Phys. Rev. B \textbf{
12}, 5105 (1975).

\bibitem{kim76} D. Kim, P. M. Levy, and J. J. Sudano, Phys. Rev.
B \textbf{13}, 2054 (1976).

\bibitem{aha77} A. Aharony, J. Phys. A \textbf{10}, 389 (1977).

\bibitem{rie81} E. K. Riedel, Physica A \textbf{106}, 110 (1981).

\bibitem{nie83} B. Nienhuis, E. K. Riedel, and M. Schick, Phys.
Rev. B \textbf{27}, 5625 (1983).

\bibitem{bax} R. Baxter \textit{Exactly Solved Models in Statistical
Mechanics} (Academic Press, New York 1982).


\bibitem{shu} See, for example, H. G. Schuster \textit{Deterministic
Chaos} (Weinheim: Physik, 1984).

\bibitem{egg74} T. P. Eggarter, Phys. Rev. Lett. \textbf{9}, 2989
(1974).

\bibitem{mul74} E. M\"{u}ller--Hartmann and J. Zittartz, Phys.
Rev. Lett. \textbf{33}, 893 (1974); Z. Phys. B \textbf{22}, 59
(1975); E. M\"{u}ller--Hartmann, Z. Phys. B \textbf{27}, 161 (1977).

\bibitem{tho82} C. J. Thompson, J. Stat. Phys. \textbf{27}, 441
(1982).

\bibitem{sto03} T. Sto\v{s}i\'{c}, B. D. Sto\v{s}i\'{c} and I. P.
Fittipaldi, Physica A \textbf{320}, 443 (2003); Physica A
\textbf{355}, 346 (2005).

\bibitem{che74} M.-S. Chen, L. Onsager, J. Bonner, and J. Nagle,
J. Chem. Phys. \textbf{60}, 405 (1974).

\bibitem{mon91} J. L. Monroe, J. Stat. Phys. \textbf{65}, 255
(1991); J. Stat. Phys. \textbf{67}, 1185 (1992).

\bibitem{ana97} N. S. Ananikian, S. K. Dallakian, N. Sh. Izmailian
and K. A. Oganessyan, Fractals, \textbf{5}, 175 (1997).

\bibitem{ara03} T. A. Arakelyan, V. R. Ohanyan, L. N. Ananikyan,
N. S. Ananikian and M. Roger, Phys. Rev. B \textbf{67}, 024424
(2003).

\bibitem{guj95}P. D. Gujrati, Phys. Rev. Lett. \textbf{74}, 809
(1995).

\bibitem{van81} J. Vannimenus, Z. Phys. B \textbf{43}, 141 (1981).

\bibitem{yok85} C. S. O. Yokoi, M. J. de Oliveira, S. R. Salinas,
Phys. Rev. Lett. \textbf{54}, 163 (1985); M. H. R. Tragtenberg, C.
S. O. Yokoi, Phys. Rev. E \textbf{52}, 2187 (1995).

\bibitem{akh92} A. Z. Akheyan and N. S. Ananikian, J. Phys.
A \textbf{25}, 3111 (1992); N. S. Ananikian, R. R. Scherbakov, J.
Phys. A \textbf{27}, L887 (1994); N. S. Ananikian, S. K. Dallakian,
B. Hu, N. Sh. Izmailian and K. A. Oganessyan, Phys. Lett. A
\textbf{248}, 381 (1998).

\bibitem{guj84} P. D. Gujrati, Phys. Rev. Lett. \textbf{53}, 2453
(1984); J. Chem. Phys. \textbf{98}, 1613 (1993).

\bibitem{pap95} Vl. V. Papoyan and R. R. Scherbakov, J. Phys.
A \textbf{28}, 6099 (1995); Fractals \textbf{4}, 105 (1996).

\bibitem{whe70} J. C. Wheeler and B. Widom, J. Chem. Phys. \textbf{52}, 5334
(1970).

\bibitem{akh94} A. Z. Akheyan and N. S. Ananikian, Phys. Lett.
A \textbf{186}, 171 (1994); JETP \textbf{107}, 196 (1995).

\bibitem{erd06} A. Erdi\c{c}, O. Canko and A. Albayrak,
J. Magn. Magn. Mater. \textbf{303}, 185 (2006).

\bibitem{mon96} J. L. Monroe, J. Phys. A \textbf{29}, 5421 (1996).

\bibitem{eck05} M. Eckstein, M. Kollar, K. Byczuk, and D.
Vollhardt, Phys. Rev. B \textbf{71}, 235119 (2005).

\bibitem{ana97} N. S. Ananikian, S. K. Dallakian, B. Hu, Complex
Systems \textbf{11}, 213 (1997).

\bibitem{nha91} N.S. Ananikian, A.R. Avakian and N.S. Izmailian, Physica A \textbf{172}, 391 (1991);
A. Z. Akheyan, and N. S. Ananikian, J. Phys. A \textbf{29}, 721
(1996).

\bibitem{alb01} E. Albayrak and M. Keskin, Eur. Phys. J. B \textbf{24}, 505 (2001);
 C. F. Delale, Int. J. Mod. Phys. \textbf{33}, 1523
(1998).

\bibitem{eki05} C. Ekiz, J. Magn. Magn. Mater. \textbf{293}, 759
(2005); \textbf{293}, 913 (2005); Physica A \textbf{347}, 353
(2005); A \textbf{353}, 286 (2005).

\bibitem{ala98} A. E. Alahverdian, N. S. Ananikian and S. K.
Dallakian, Phys. Rev. E \textbf{57}, 2452 (1998); N. S. Ananikian,
R. G. Ghulghazaryan, Phys. Lett. A \textbf{277}, 249 (2000); R. G.
Ghulghazaryan, N. S. Ananikian and P. M. A. Sloot, Phys. Rev. E
\textbf{66}, 046110 (2002);  R. G. Ghulghazaryan, N.S. Ananikian,
J.Phys. A \textbf{36}, 6297 (2003).

\bibitem{wu} F. Y. Wu, Rev. Mod. Phys. \textbf{54}, 235 (1982).

\bibitem{graph} B. Balobas, \textit{Modern Graph Theory} (Springer, New York,
1998).



\end{thebibliography}
\end{document}